\documentclass[aps,twocolumn,english,superscriptaddress,citeautoscript,preprintnumbers,amsmath,amssymb,floatfix,footinbib]{revtex4-1}
\usepackage{hyperref}
\hypersetup{colorlinks=true,linkcolor=red,citecolor=blue}
\usepackage{amsmath}
\usepackage{graphicx}
\usepackage{dcolumn}
\usepackage{float}

\begin{document}
\title{TaCo$_{2}$Te$_{2}$: An air-stable, magnetic van der Waals material with high mobility}

\author{Ratnadwip Singha}
    \affiliation{Department of Chemistry, Princeton University, Princeton, New Jersey 08544, USA}

\author{Fang Yuan}
    \affiliation{Department of Chemistry, Princeton University, Princeton, New Jersey 08544, USA}

\author{Guangming Cheng}
    \affiliation{Princeton Institute for Science and Technology of Materials, Princeton University, Princeton, New Jersey 08544, USA}

\author{Tyger H. Salters}
    \affiliation{Department of Chemistry, Princeton University, Princeton, New Jersey 08544, USA}

\author{Yuzki M. Oey}
    \affiliation{Materials Department and Materials Research Laboratory, University of California, Santa Barbara, California 93106, USA}

\author{Graciela V. Villalpando}
    \affiliation{Department of Chemistry, Princeton University, Princeton, New Jersey 08544, USA}

\author{Milena Jovanovic}
    \affiliation{Department of Chemistry, Princeton University, Princeton, New Jersey 08544, USA}

\author{Nan Yao}
    \affiliation{Princeton Institute for Science and Technology of Materials, Princeton University, Princeton, New Jersey 08544, USA}

\author{Leslie M. Schoop}
    \affiliation{Department of Chemistry, Princeton University, Princeton, New Jersey 08544, USA}

\begin{abstract}

Van der Waals (vdW) materials are an indispensable part of functional device technology due to their versatile physical properties and ease of exfoliating to the low-dimensional limit. Among all the compounds investigated so far, the search for magnetic vdW materials has intensified in recent years, fueled by the realization of magnetism in two dimensions (2D). However, \textit{metallic} magnetic vdW systems are still uncommon. In addition, they rarely host high-mobility charge carriers, which is an essential requirement for high-speed electronic applications. Another shortcoming of 2D magnets is that they are highly air sensitive. Using chemical reasoning, we introduce TaCo$_{2}$Te$_{2}$ as an air-stable, high-mobility, magnetic vdW material. It has a layered structure, which consists of Peierls distorted Co chains and a large vdW gap between the layers. We find that the bulk crystals can be easily exfoliated and the obtained thin flakes are robust to ambient conditions after four months of monitoring using an optical microscope. We also observe signatures of canted antiferromagntic behavior at low-temperature. TaCo$_{2}$Te$_{2}$ shows a metallic character and a large, non-saturating, anisotropic magnetoresistance. Furthermore, our Hall data and quantum oscillation measurements reveal the presence of both electron- and hole-type carriers and their high mobility.

\end{abstract}

\maketitle

\section{Introduction}

Van der Waals (vdW) materials are the important building blocks of low-dimensional device technology as they provide the flexibility of designing a large number of unique heterostructures by stacking two-dimensional (2D) atomic layers \cite{Novoselov1}. These layers can be chosen from an expansive catalogue of predicted vdW compounds \cite{Mounet} with different experimentally verified functionalities ranging from superconductivity \cite{Xi,Qi} and metallic or semiconducting transport \cite{Han,Wang5}, to optical properties \cite{Zhou,Gao} and magnetism \cite{Burch}. Such assembled structures offer capabilities to manipulate the properties of different layers, for example, the proximity effect often leads to new quantum states at the interfaces \cite{Kezilebieke,Tang}.

Out of all the vdW systems studied so far, magnetic vdW compounds have been of particular interest due to the discovery of the long-sought 2D magnetism and its high tunability \cite{Huang2,Huang3,Gong,Zhang,Wang,Ghazaryan,Cai,Weber,Lee}. 2D magnets have various technological applications, particularly in magnetic tunnel junctions \cite{Song,Lin}. In spite of such intense demand, the examples of true magnetic vdW materials (i.e. exfoliable down to few atomic layers) are rather limited. Moreover, almost all of them show either semiconducting or insulating electronic transport properties. Exceptions include metallic ferromagnets Fe$_{3}$GeTe$_{2}$ \cite{Wang2}, Fe$_{4}$GeTe$_{2}$ \cite{Seo} and antiferromagnets $R$Te$_{3}$ ($R$=rare earth) \cite{Ru1,Ru2,Lei,Dalgaard}. Among these, GdTe$_{3}$ has been highlighted for its high-mobility charge carriers that can play an important role in ultrafast, low-power electronics \cite{Liao}. The only comparable high-mobility vdW systems are non-magnetic graphite \cite{Dillon} and black phosphorus \cite{Akahama}, both of which have demonstrated their versatility in device fabrication \cite{Novoselov2,Kostarelos,Li1,Gomez,Li2}. Of course, graphite has shown many more interesting features especially generated by the Moir\'{e} pattern in twisted multilayer graphene \cite{Cao1,Cao2}, including a novel ferromagnetic state in twisted bilayer graphene \cite{Sharpe}. With intrinsic magnetism as another controllable parameter, GdTe$_{3}$ is a prime material candidate for spintronics applications \cite{Bhatti} or creating twisted multilayer devices with magnetic order and metallic conductivity. However, one major limitation of this compound (and also the $R$Te$_{3}$ family in general) is its robustness, as the exfoliated GdTe$_{3}$ flakes degrade quickly after a brief exposure to air \cite{Lei}. This introduces additional complexities during handling such as inert atmosphere exfoliation and encapsulation, as well as hinders its wide-scale use. An air-stable, high-mobility, magnetic vdW compound is therefore highly desired.

\begin{figure*}
\includegraphics[width=0.9\textwidth]{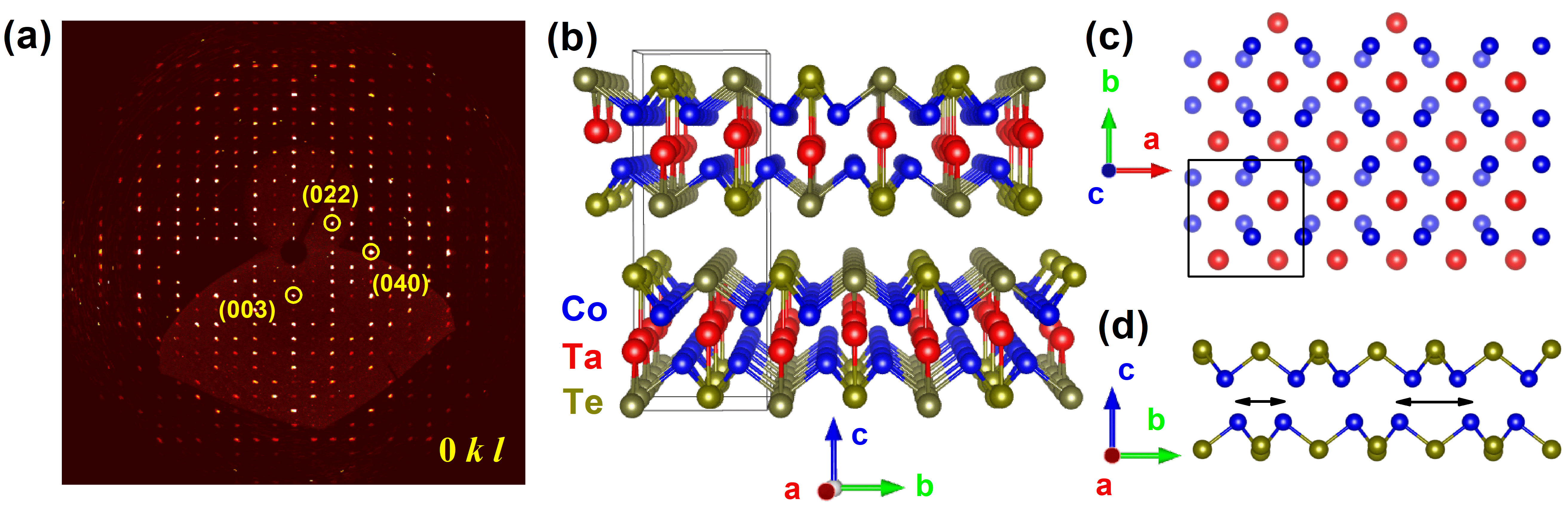}
\caption{Crystal structure of TaCo$_{2}$Te$_{2}$. a) Indexed precession single crystal x-ray diffraction image along the [0$kl$] direction of a TaCo$_{2}$Te$_{2}$ single crystal. b) The solved crystal structure. c) Each layer consists of two interpenetrating square-nets of Ta atoms and Co dimers in the $ab$-plane. d) The distorted Co chains along the $a$-axis. The black arrows illustrate the changing distance between two consecutive Co atoms within a chain.}
\label{fig:boat1}
\end{figure*}

To design such a material, it helps to first understand the other high-mobility vdW magnet, GdTe$_{3}$. Its structure contains Te square-nets that are susceptible to a Peierls distortion, which results in the formation of a charge density wave (CDW). The square-net motif has previously been linked to high electronic mobility, as it contains delocalized, hypervalent bonds \cite{Klemenz}. Electron delocalization is often in competition with a Peierls distortion or a CDW; the distortion (partially) localizes these. Nonetheless, the partial delocalization can remain after the distortion, which is why the presence of a Peierls instability can be a good indicator for high mobility charge carriers \cite{Khoury}. Black phosphorus, for example has a distorted structure \cite{Gomez2}, whereas in graphene such distortion is prevented by the delocalized electrons \cite{Schoop}. Using this chemical intuition, we searched for a layered material that exhibits a Peierls distortion in addition to magnetic species. TaCo$_{2}$Te$_{2}$ had been reported in a distorted layered structure and contains Co, thus fulfilling all the chemical requirements. We here show that it indeed features magnetism, high mobility, and can be exfoliated. Optical microscopy images show that the exfoliated flakes are stable in air for months.

\section{Results and Discussions}

\subsection{Crystal structure and exfoliation of the bulk sample}

The scanning electron microscope (SEM) image of a TaCo$_{2}$Te$_{2}$ single crystal is shown in \textbf{Figure S1} (see the supporting information). The energy dispersive x-ray spectroscopy (EDX) analysis on different regions of the crystal reveals almost ideal stoichiometry (\textbf{Figure S2} in the supporting information). Single crystal x-ray diffraction (SCXRD) confirms the previously reported structure \cite{Tremel1}. \textbf{Figure 1}a displays a typical indexed precession diffraction image. All the parameters related to the structural solution are summarized in \textbf{Table S1}, \textbf{S2}, and \textbf{S3} in the supporting information. The corresponding structure is presented in Figure 1b. TaCo$_{2}$Te$_{2}$ crystallizes in a layered orthorhombic structure with spacegroup $Pmcn$ (62). The structure consists of quintuple layers that are stacked along the crystallographic $c$-axis and only held together by weak vdW forces (Figure 1b). Each quintuple layer consists of a puckered square-net layer of Ta atoms (Figure 1c) in the center, sandwiched by square-net layers of Co dimers, which again are sandwiched by Te layers. The Co dimers are tilted alternately with respect to the Ta square-net (Figure 1c). The Te atoms occupy the spaces in between the Ta and Co square-nets. The structure is distorted in respect to similar compounds such as TaNi$_{2}$Te$_{2}$ \cite{Tremel2}; the distortion becomes apparent if viewed along the $a$-axis. When the Ta atoms are ignored for a moment, one can see that the quintuple layer contains a pair of cobalt-tellurium zigzag chains, where the Co-Co distance alternates within every chain (Figure 1d). This Peierls distortion appears due to the structural instability in the quasi one-dimensional zigzag chains. Peierls distortions often lead to the formation of a CDW in a material \cite{Pouget}. In TaCo$_{2}$Te$_{2}$, however, we have not observed any indications of a higher symmetry structure in SCXRD experiments up to 390 K.

\begin{figure*}
\includegraphics[width=0.8\textwidth]{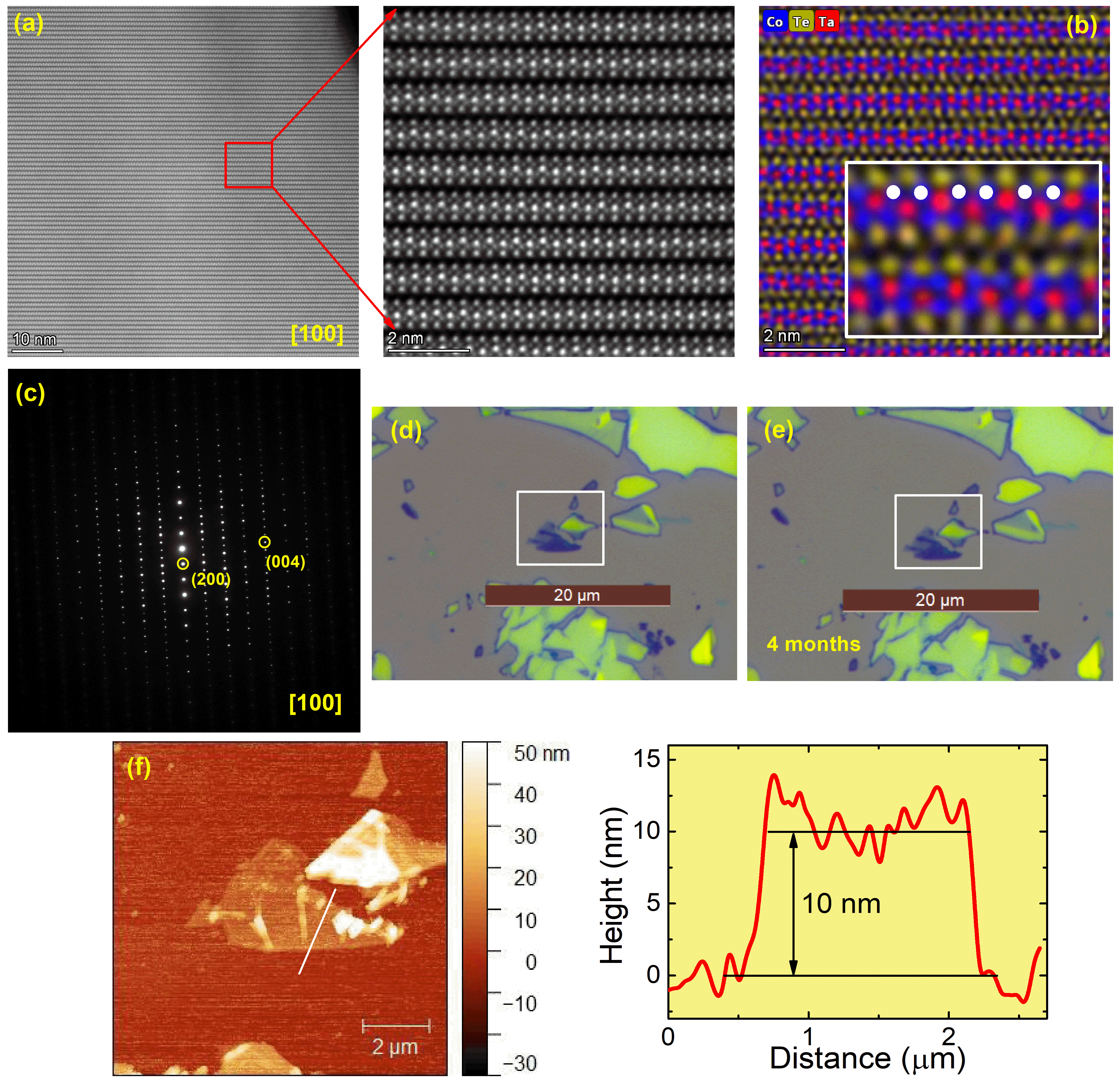}
\caption{Atomic scale imaging and mechanical exfoliation of TaCo$_{2}$Te$_{2}$ single crystals. a) Extreme high-resolution, high-angle annular dark-field scanning transmission electron microscopy image of a crystal along the $a$-axis, showing the van der Waals gap between the layers. b) Elemental mapping of the image. The inset shows an enlarged view where subsequent Co atoms are highlighted by white points to emphasize the Peierls-distorted structure. c) Selected area electron diffraction pattern along the [100] axis. A typical TaCo$_{2}$Te$_{2}$ thin flake d) immediately after exfoliation and e) after exposure to air for four months. f) Atomic force microscopy image of the thin flake (left panel) and the obtained height profile (right panel).}
\label{fig:boat1}
\end{figure*}

While the vdW gap in the crystal structure suggests that exfoliation of this material should be possible, in reality, exfoliation often can be hindered by defects that occupy the gap \cite{Purbawati}. In addition, while TaCo$_{2}$Te$_{2}$ is listed in the ICSD, it does not appear as an exfoliable material in the 2D materials database \cite{Mounet}, where materials were screened computationally for their exfoliation capability. To elucidate whether the vdW gap is large and free of defects, we studied the crystals with atomic-resolution high-angle annular dark-field scanning transmission electron microscopy (HAADF-STEM). In \textbf{Figure 2}a, the HAADF-STEM image of a TaCo$_{2}$Te$_{2}$ crystal viewed along the $a$-axis clearly demonstrates the large vdW gap between the layers. Using the images, we could calculate the vdW gap to be $\sim$0.314 nm. This value is higher than that for graphite and $h$-BN, as well as for the extensively studied transition metal dichalcogenides \cite{Liu}. Thus, it should be possible to exfoliate the TaCo$_{2}$Te$_{2}$ crystals down to very thin flakes or even monolayers. Remarkably, the atomic-resolution elemental mapping, shown in Figure 2b, directly visualizes the Peierls distortion along the Co chains. The inset of Figure 2b shows an enlarged view with the consecutive Co atoms highlighted by white points. We determined that the distance changes from $\sim$2.82 to $\sim$4.02 {\AA} between alternate Co-Co atomic pairs within a chain. These values are in agreement with the x-ray diffraction data [2.552(4) and 4.019(4) {\AA}]. The obtained selected area electron diffraction (SAED) pattern in Figure 2c matches well with the simulated pattern from the reported crystal structure and confirms the high crystalline quality of the grown TaCo$_{2}$Te$_{2}$ samples.

To investigate the viability of using TaCo$_{2}$Te$_{2}$ in low-dimensional devices, we exfoliated the single crystals mechanically using Scotch tape and transferred the flakes on a silicon wafer. In Figure 2d an optical image of a typical thin flake is highlighted by the white box. This thin flake looks visually unaffected under the optical microscope after exposure to air for at least four months (Figure 2e). Nevertheless, we can not exclude the possibility of microstructural changes due to mild air sensitivity. This is in stark contrast to GdTe$_3$, where fast degradation can be observed under the optical microscope \cite{Lei}. The demonstrated air stability is remarkable for the sheets, especially for a telluride, as these tend to be very air sensitive \cite{Ye,Mirabelli}. Therefore, TaCo$_{2}$Te$_{2}$ is an ideal material for device fabrication. In order to determine the thickness of the flake, we performed atomic force microscopy (AFM), as shown in Figure 2f (left panel). The thinnest part of the flake is $\sim$10 nm (Figure 2f: right panel). This thickness corresponds to about 11 layers.

\begin{figure*}
\includegraphics[width=0.6\textwidth]{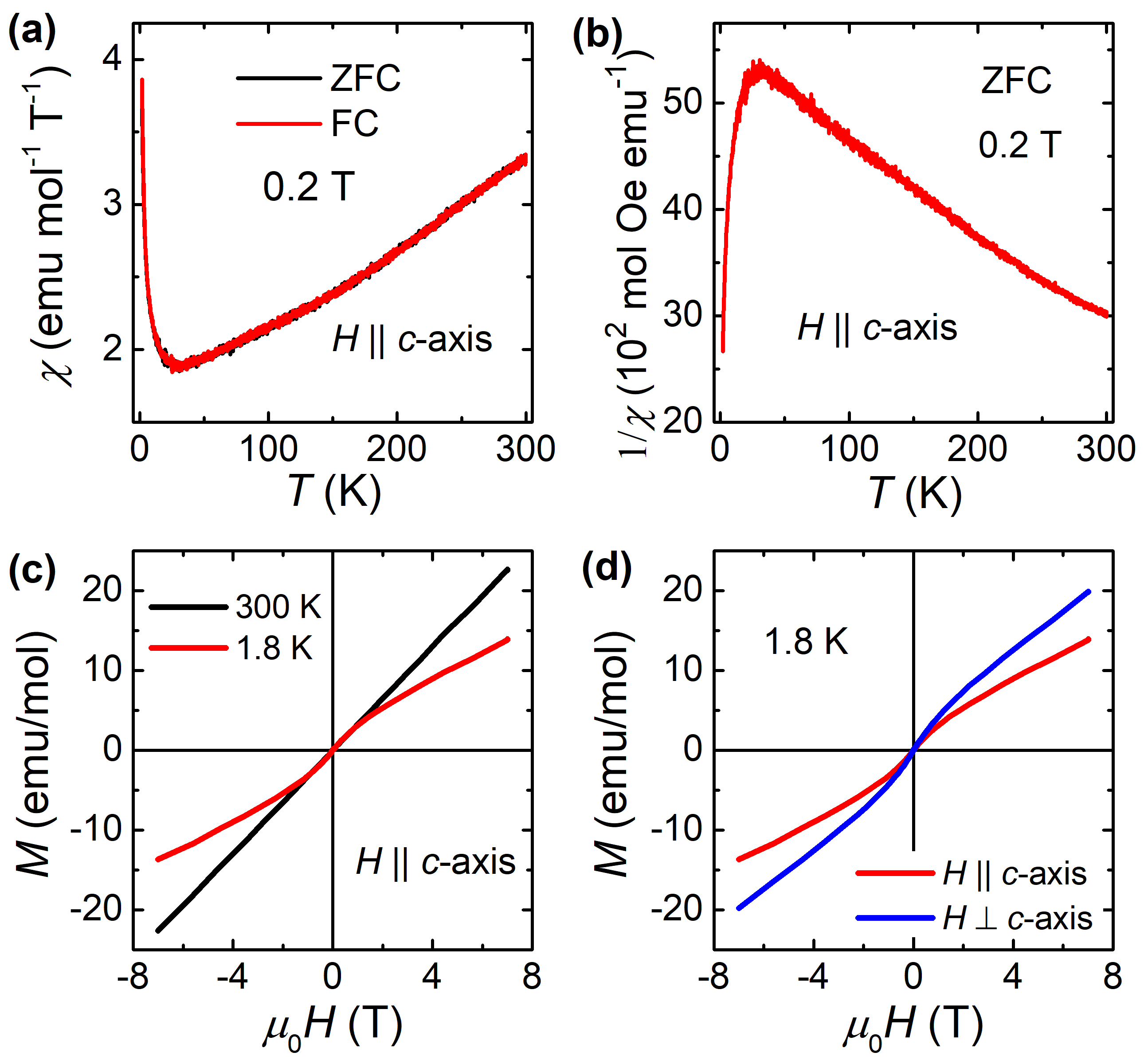}
\caption{Magnetic properties of the TaCo$_{2}$Te$_{2}$ crystals. a) Temperature dependence of the magnetic susceptibility ($\chi$) under zero-field-cooled (ZFC) and field-cooled (FC) conditions with the field applied along the $c$-axis. b) Inverse magnetic susceptibility (1/$\chi$) plotted as a function of temperature for the magnetic field applied along the $c$-axis. c) Magnetic field ($H$) dependence of the magnetization ($M$). d) $M$($H$) curves at 1.8 K for magnetic field applied along the $c$-axis and along the $ab$-plane.}
\label{fig:boat1}
\end{figure*}

\subsection{Magnetic properties of the bulk single crystals}

Next, we study the magnetic properties of the TaCo$_{2}$Te$_{2}$ crystals. The temperature ($T$) dependence of the magnetic susceptibility ($\chi$) is shown in \textbf{Figure 3}a in both zero-field-cooled (ZFC) and field-cooled (FC) conditions with the field applied along the $c$-axis. The experimental data reveal a rather unusual behavior, where $\chi$ ($T$) decreases monotonically from room temperature followed by a sharp increase below $\sim$14 K without any significant bifurcation between ZFC and FC curves. This results in a 1/$\chi$-plot that is drastically different from that what would be expected for Curie-Weiss behavior (Figure 3b). While we could not identify a clear susceptibility peak corresponding to long-range magnetic ordering down to 1.8 K, the unusual $\chi$ vs. $T$ curve might suggest that magnetic order or fluctuations exist at room temperature. We find that similar magnetic properties have recently been reported for a fractional valence state iridate La$_{3}$Ir$_{3}$O$_{11}$, where strong geometrical frustration prevents long-range magnetic ordering, making it a potential candidate for observing the spin-liquid state \cite{Yang}. This frustrated structure and competing magnetic exchange interactions within the lattice also lead to non-Curie-Weiss paramagnetism at high temperature \cite{Yang}, which is also the case for TaCo$_{2}$Te$_{2}$ at least up to 300 K (Figure 3b). If indeed TaCo$_{2}$Te$_{2}$ has a frustrated magnetic structure, which is not unexpected given the complex arrangement of the Co atoms in each layer, it would be interesting to probe the magnetism with changing layer numbers and in the monolayer limit. It is possible that TaCo$_{2}$Te$_{2}$ might reach the pure paramagnetic state well above the room temperature. The magnetic field ($H$) dependence of the magnetization ($M$) in Figure 3c shows a canted antiferromagnetic behavior at 1.8 K without any signature of saturation up to 7 T. The  nature of the $M$($H$) curve remains similar when the field is applied along the $ab$-plane (Figure 3d). We also observe a clear magnetocrystalline anisotropy between these two crystallographic directions. On the other hand, the $M$($H$) curve resembles either an antiferromagnet or paramagnet at room temperature (Figure 3c).

\subsection{Electronic transport properties}

Having established TaCo$_{2}$Te$_{2}$ as an air-stable, antiferromagnetic, exfoliable material, we now turn to the electronic transport properties of bulk TaCo$_{2}$Te$_{2}$ crystals. \textbf{Figure 4}a displays the the temperature dependent resistivity ($\rho_{xx}$) with the current along the $a$-axis; a metallic character is observed throughout the measured temperature range, i.e. it decreases monotonically from room temperature. The residual resistivity ratio [RRR=$\rho_{xx}$(300 K)/$\rho_{xx}$(2 K)] is $\sim$17, pointing to decent crystal quality. In the low-temperature region below 25 K, $\rho_{xx}$ follows a $T^{4}$-behavior (Figure 4a: inset), which is distinct from a quadratic temperature dependence expected from Fermi liquid theory (pure electron-electron scattering), or a $T^{5}$-dependence expected for the electron-phonon scattering \cite{Ziman}. We note that $T^{4}$-dependence could originate from interband electron-phonon scattering \cite{Ziman,Singha}.

Upon applying a magnetic field along the $c$-axis, a large, non-saturating magnetoresistance \newline
[MR=$\frac{\rho_{xx}(\mu_{0}H)-\rho_{xx}(0)}{\rho_{xx}(0)}\times$100 \%] is observed (Figure 4b). At 2 K and 9 T, the MR reaches $\sim$263 \% and decreases with increasing temperature. The origin of such large MR is much debated, especially in the context of topological semimetals. However, one possible explanation is provided by the semiclassical electron theory for a material with compensated density of the electron- and hole-type carriers \cite{Ziman}. In this scenario, the MR should show a quadratic magnetic field dependence. In TaCo$_{2}$Te$_{2}$, the MR follows a $H^{1.5}$-type relation, which is close but not identical to the expected $H^{2}$-dependence for a fully-compensated system. We also measured the directional dependence of this MR by rotating the magnetic field in the $bc$-plane while keeping the current direction along the $a$-axis (Figure 4c: schematic). This measurement configuration ensures that the current is always perpendicular to the applied magnetic field direction and there is no contribution of the changing Lorentz force in the directional dependence of the MR. The two-fold symmetric pattern of the polar plot in Figure 4c reveals the highly anisotropic nature of the Fermi surface, which is expected given the quasi-two-dimensional character of the TaCo$_{2}$Te$_{2}$ crystals. Furthermore, several additional weak kinks are observed at angles 17$^{\circ}$, 32$^{\circ}$, 47$^{\circ}$, and 63$^{\circ}$ (and symmetrically in all of the four quadrants), indicating a complex structure of the Fermi surface.

\begin{figure*}
\includegraphics[width=0.8\textwidth]{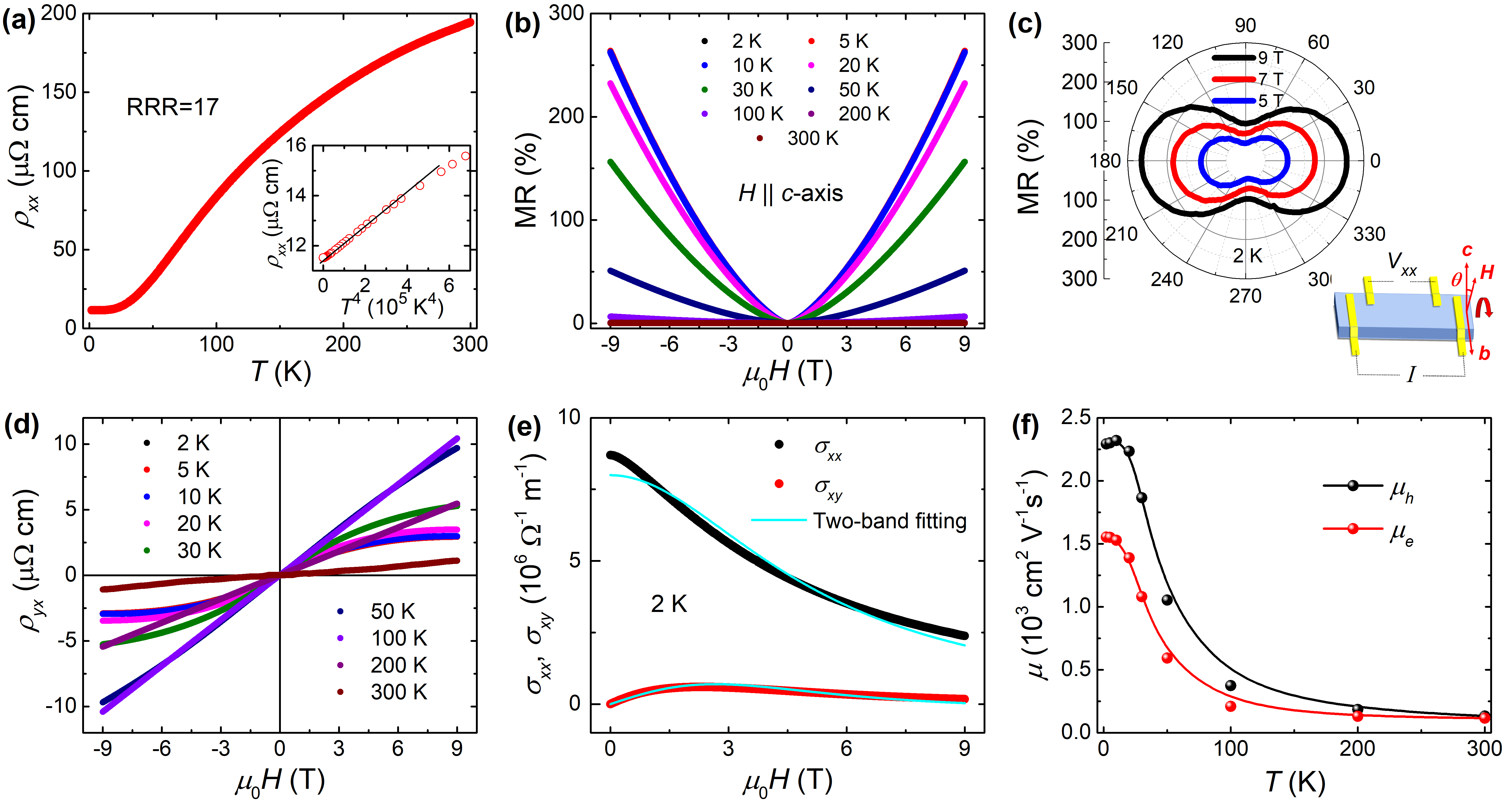}
\caption{Electronic transport properties of TaCo$_{2}$Te$_{2}$ single crystals. a) Temperature ($T$) dependence of the resistivity ($\rho_{xx}$) with the current along the $a$-axis. The inset shows the $\rho_{xx}$$\propto$$T^{4}$-type behavior in the low-temperature region. b) Magnetoresistance (MR) at different temperatures with the magnetic field applied along the $c$-axis. c) Directional dependence of the MR at 2 K, when the magnetic field is rotated within the $bc$-plane. Schematic illustrating the experimental configuration. d) Magnetic field dependence of the Hall resistivity ($\rho_{yx}$) at different temperatures. e) Simultaneous fitting of the longitudinal conductivity ($\sigma_{xx}$) and Hall conductivity ($\sigma_{xy}$) using the two-band model. f) Temperature dependence of the mobility for electron- ($\mu_{e}$) and hole-type ($\mu_{h}$) carriers. The lines are guide to the eye.}
\label{fig:boat1}
\end{figure*}

To gain insight into the nature of the charge carriers, Hall effect measurements were performed. Figure 4d shows the Hall resistivity ($\rho_{yx}$) as a function of the magnetic field at different temperatures. The non-linear field dependence of the $\rho_{yx}$($H$) curves confirms the presence of multiple Fermi pockets in TaCo$_{2}$Te$_{2}$. To analyze the experimental data, we use the following two-band model, which takes into account the contributions from both electron and hole-type carriers \cite{Hurd},

\begin{equation}
\sigma_{xy}=\left[\frac{n_{h}\mu_{h}^{2}}{1+(\mu_{h}\mu_{0}H)^2}-\frac{n_{e}\mu_{e}^{2}}{1+(\mu_{e}\mu_{0}H)^2}\right]e\mu_{0}H,
\end{equation}
\begin{equation}
\sigma_{xx}=e\left[\frac{n_{h}\mu_{h}}{1+(\mu_{h}\mu_{0}H)^{2}}+\frac{n_{e}\mu_{e}}{1+(\mu_{e}\mu_{0}H)^{2}}\right],
\end{equation}
where $\sigma_{xy}$=$\frac{\rho_{yx}}{\rho_{yx}^{2}+\rho_{xx}^{2}}$ and $\sigma_{xx}$=$\frac{\rho_{xx}}{\rho_{yx}^{2}+\rho_{xx}^{2}}$ are Hall conductivity and longitudinal conductivity, respectively. $n_{h}$ ($n_{e}$) and $\mu_{h}$ ($\mu_{e}$) are hole (electron) density and mobility, respectively. Using the Equations 1 and 2, the $\sigma_{xy}$($H$) and $\sigma_{xx}$($H$) curves are fitted simultaneously for different temperatures. In Figure 4e, one such fitting is shown for a representative temperature. From the extracted fitting parameters, we find the electron and hole density to be 1.442(9)$\times$10$^{20}$ and 1.202(6)$\times$10$^{20}$ cm$^{-3}$, respectively. These values reveal that TaCo$_{2}$Te$_{2}$ is a nearly-compensated semimetal, which is consistent with the nature of the MR curves as already discussed. At 2 K, the obtained electron and hole mobilities are \newline
1.553(9)$\times$10$^{3}$ and 2.293(11)$\times$10$^{3}$ cm$^{2}$V$^{-1}$s$^{-1}$, respectively. Although these values are one order of magnitude smaller than the highest quality GdTe$_{3}$ crystals \cite{Lei}, they are still very high and comparable to several topological semimetals \cite{Novak,Singha2}. In Table 1, we have compared the carrier mobility for TaCo$_{2}$Te$_{2}$ with that for different magnetic layered compounds. It is evident that this compound is a rare air-stable, magnetic vdW system which also hosts high-mobility charge carriers. As Figure 4f illustrates, the mobility for both types of carriers decreases with increasing temperature.

\begin{center}
\begin{table*}
\caption{Comparison of the carrier mobility in bulk single crystals of metallic, magnetic layered compounds. We have also listed the magnetic state and whether it is an air-stable van der Waals (vdW) material.}
 \begin{tabular}{|c c c c c c c c c c c c c|}
 \hline
  & & & & & & & & & & & &\\
  & Material & & Magnetic state & & vdW? & & Air-stable? & & Mobility & & Reference &\\

  & & & & & & & & & cm$^{2}$V$^{-1}$s$^{-1}$ & & &\\ [0.5ex]
 \hline\hline
  & & & & & & & & & & & &\\

  & GdTe$_{3}$ & & Antiferromagnetic & & \textbf{Yes} & & No & & 23500-61200 & & \cite{Lei} &\\

  & & & & & & & & & & & &\\

  & NdTe$_{3}$ & & Antiferromagnetic & & \textbf{Yes} & & No & & 60000 & & \cite{Dalgaard} &\\

  & & & & & & & & & & & &\\

  & EuMnBi$_{2}$ & & Antiferromagnetic & & No & & Unknown & & 14000 & & \cite{Masuda} &\\

  & & & & & & & & & & & &\\

  & TbTe$_{3}$ & & Antiferromagnetic & & \textbf{Yes} & & No & & 2000-8000 & & \cite{Xing} &\\

  & & & & & & & & & & & &\\

  & YbMnSb$_{2}$ & & Antiferromagnetic & & No & & \textbf{Yes} & & 1310-6538 & & \cite{Wang3} &\\

  & & & & & & & & & & & &\\

  & TaCo$_{2}$Te$_{2}$ & & Canted antiferromagnetic & & \textbf{Yes} & & \textbf{Yes} & & 1553-2293 & & This work &\\

  & & & & & & & & & & & &\\

  & BaMnSb$_{2}$ & & Antiferromagnetic & & No & & Unknown & & 1300 & & \cite{Liu2} &\\

  & & & & & & & & & & & &\\

  & BaFe$_{2}$As$_{2}$ & & Antiferromagnetic & & No & & \textbf{Yes} & & 1130 & & \cite{Kuo} &\\

  & & & & & & & & & & & &\\

  & Sr$_{1-y}$Mn$_{1-z}$Sb$_{2}$ & & Ferromagnetic & & No & & No & & 570 & & \cite{Liu3} &\\

  & & & & & & & & & & & &\\

  & CaMnBi$_{2}$ & & Antiferromagnetic & & No & & Unknown & & 488 & & \cite{Wang4} &\\

  & & & & & & & & & & & &\\

  & SrMnBi$_{2}$ & & Antiferromagnetic & & No & & No & & 250 & & \cite{Park} &\\

  & & & & & & & & & & & &\\

  & Co$_{3}$Sn$_{2}$S$_{2}$ & & Ferromagnetic & & No & & \textbf{Yes} & & $\sim$100 & & \cite{Shen} &\\

  & & & & & & & & & & & &\\

  & Fe$_{3}$GeTe$_{2}$ & & Ferromagnetic & & \textbf{Yes} & & \textbf{Yes} & & $\sim$2.8 & & \cite{Wang2} &\\

  & & & & & & & & & & & &\\
 \hline
\end{tabular}
\end{table*}
\end{center}

\begin{figure}
\includegraphics[width=0.45\textwidth]{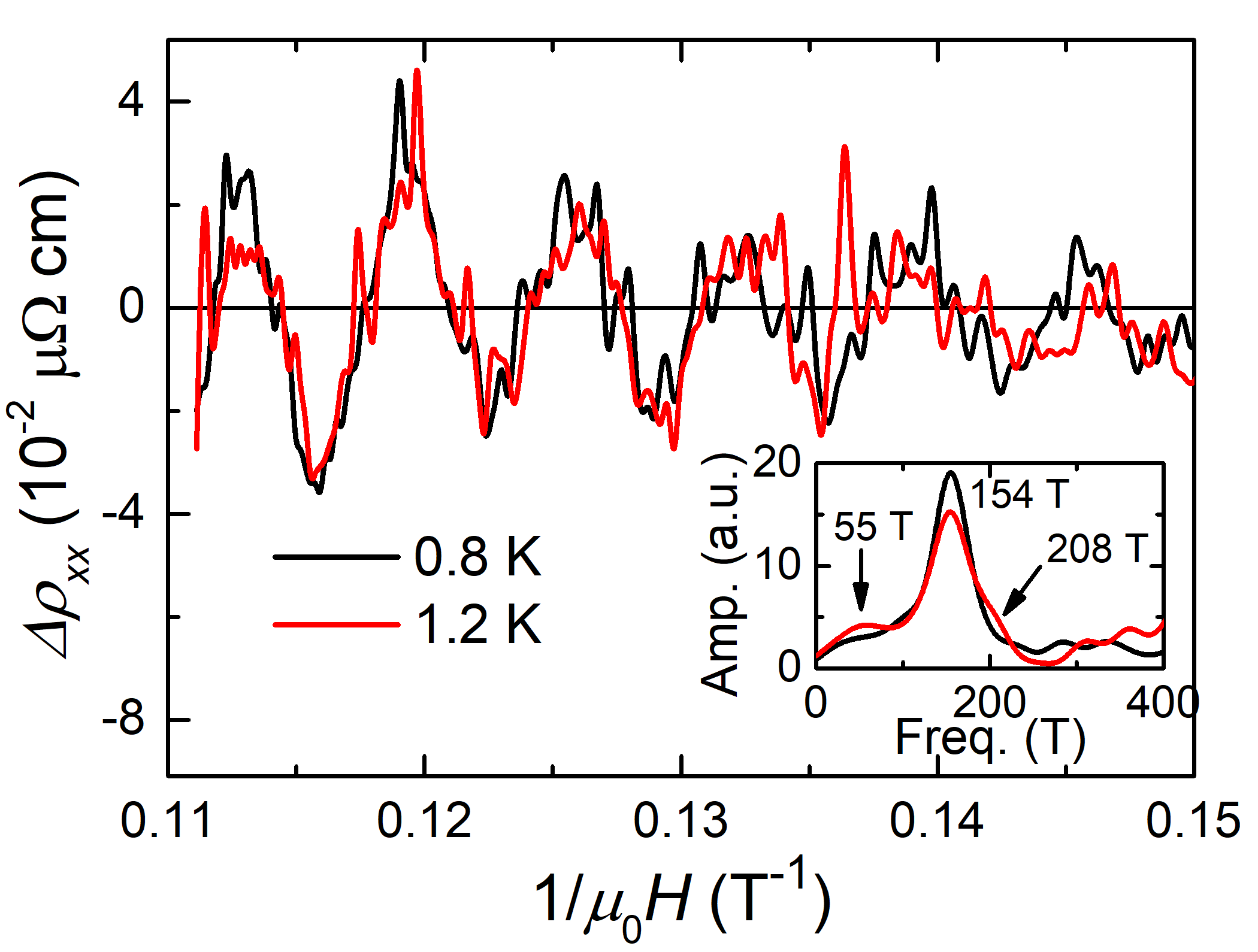}
\caption{Shubnikov-de Haas quantum oscillations in the TaCo$_{2}$Te$_{2}$ crystals. The inset shows the corresponding fast Fourier transform spectrum.}
\label{fig:boat1}
\end{figure}

As the TaCo$_{2}$Te$_{2}$ crystals are cooled below 2 K, we observe prominent Shubnikov-de Haas (SdH) quantum oscillations in the MR for magnetic field along the $c$-axis. The appearance of the SdH oscillations also suggests a long mean free path and hence high mobility of the charge carriers. After subtracting a smooth background from the experimental data, the extracted oscillatory part ($\Delta\rho_{xx}$) of the resistivity is plotted as a function of the inverse magnetic field in \textbf{Figure 5} for two different temperatures. The corresponding fast Fourier transform (FFT) spectrum (shown in the inset) reveals three oscillation frequencies ($F$), 55, 154, and 208 T, thus further confirming the presence of multiple Fermi pockets in TaCo$_{2}$Te$_{2}$. From these quantum oscillation frequencies, we can calculate the extremum cross-sectional area ($A_{F}$) of the Fermi surface perpendicular to the magnetic field direction by using the Onsager relation, $F=(\phi_{0}/2\pi^{2})A_{F}$, where $\phi_{0}$ is the magnetic flux quantum. The obtained cross-sectional areas are 5.24$\times$10$^{-3}$, 14.68$\times$10$^{-3}$, and 19.82$\times$10$^{-3}$ {\AA}$^{-2}$, respectively.

\section{Outlook and Conclusions}

We demonstrate that chemical principles can be used to find 2D materials with rare combination of physical properties. Our strategy, based on the presence of a structural distortion in combination with a layered structure and the presence of typically magnetic elements, has proven fruitful for finding a new, air-stable, high-mobility magnetic vdW material. It is worth noting that our chemical reasoning led us to a 2D material that was overlooked in extended computational searches, as TaCo$_{2}$Te$_{2}$ does not appear in the 2D materials database \cite{Mounet}. The results of our band structure calculations (\textbf{Figure S3}a) for bulk TaCo$_{2}$Te$_{2}$ reveal that the Fermi surface of this material is very complex as a large number of bands reside at the Fermi energy. Thus this material might have been considered unappealing due to the messy band structure. The steep nature of the bands, however, could give an indication of the high-mobility carriers. Our theoretical calculations further predict that the metallic character of TaCo$_{2}$Te$_{2}$ persists down to the monolayer limit (Figure S3b).

To conclude, we showed that the single-crystalline vdW compound TaCo$_{2}$Te$_{2}$ is a great candidate for 2D device fabrication. This material has large vdW gap and hence the bulk crystals can easily be exfoliated to few layers. Moreover, the thin flakes are robust in air for at least four months as observed in optical microscopy. Magnetization measurements show canted antiferromagnetic behavior at low temperature with possible magnetic frustration. TaCo$_{2}$Te$_{2}$ is metallic and shows a large, non-saturating, anisotropic MR, which possibly originates from the nearly compensated electron and hole density in this material. Hall effect measurements reveal high mobility for both types of charge carriers. We have observed quantum oscillations in the MR data, which also confirm the presence of multiple Fermi pockets. Thus, TaCo$_{2}$Te$_{2}$ is a rare magnetic vdW compound, which offers a range of unique functionalities highly desired for low-dimensional device technology.\\

\section{Experimental and theoretical details}

\textbf{Single crystal growth and determination of the stoichiometry:} Single crystals of TaCo$_{2}$Te$_{2}$ were grown with chemical vapor transport using iodine as transport agent. First, the polycrystalline powder was prepared in a solid state reaction. Elemental Ta (Sigma-Aldrich 99.99\%), Co (Sigma-Aldrich 99.9\%), Te (Sigma-Aldrich 99.999\%) were mixed in molar ratio and sealed in a evacuated quartz tube, which was then heated to 900 $^{\circ}$C for 3 days. Next, the resulting powder was mixed with iodine (Sigma-Aldrich 99.99\%) and sealed in another quartz tube under vacuum. The tube was placed in a gradient furnace for 7 days keeping the hotter end with the powder at 950 $^{\circ}$C and the other end of the tube at 850 $^{\circ}$C. After cooling down, plate-like crystals were mechanically extracted from the colder end of the quartz tube. The stoichiometry of the grown crystals was checked using EDX in a Verios 460 SEM operating at 15 kV and equipped with an Oxford EDX detector.

\textbf{X-ray diffraction measurements:} The SCXRD measurements were performed using a Bruker D8 VENTURE diffractometer fitted with a PHOTON III CPAD detector and graphite-monochromatized Mo-$K_{\alpha}$ radiation source. Raw data were corrected for background, polarization, Lorentz factors, and multiscan absorption. Structure refinements were performed using the OLEX2 software package.

\textbf{Scanning/transmission electron microscopy:} TEM thin lamella for microstructure characterization were prepared by focused ion beam cutting via a FEI Helios NanoLab$^{TM}$ 600 dual beam system (FIB/SEM). Conventional TEM imaging, atomic resolution HAADF-STEM imaging, and atomic-level EDX mapping were performed on a double Cs-corrected FEI Titan Cubed Themis 300 scanning/transmission Electron Microscope (S/TEM) equipped with an X-FEG source operated at 300 kV and Super-X EDX system. The microscope is equipped with a Gatan Quantum SE/963 P post-column energy filter for energy filtered TEM. The high-resolution scanning transmission electron microscopy (HRSTEM) images were processed using the Gatan Microscopy Suite (GMS 3.4.3) software. The electron diffraction simulations were performed using the CrystalMaker and SingleCrystal software package.

\textbf{Atomic force microscopy:} The AFM of the thin flakes was performed using a Bruker NanoMan AFM. The raw data were processed by Gwyddion software package.

\textbf{Electronic transport and magnetization measurements:} The electronic transport measurements of the TaCo$_{2}$Te$_{2}$ crystals were done in a physical property measurement system (Quantum Design) using the ac-transport option. Electrical contacts were attached on the crystals in four-probe configuration using gold wires and silver paste. For achieving sample temperatures lower than 1.8 K, a dilution refrigerator insert was used. The magnetization data were collected in a SQUID-VSM MPMS3 (Quantum Design).

\textbf{Band structure calculations:} Band structure calculations were carried out using VASP 5.4.4. software \cite{kresse1996efficiency, kresse1996efficient}. Geometry of bulk TaCo$_{2}$Te$_{2}$ was taken from the crystallographic data, while a monolayer was created by increasing the distance between layers from 3.8782 to 24.2748 {\AA}. Geometries were not relaxed. Band structures  were calculated using the Strongly Constrained and Appropriately Normed Semilocal Density Functional (SCAN) \cite{sun2015strongl} for exchange and correlation, and Ta\_pv, Te, and Co Projector Augmented Wave (PAW) potentials \cite{blochl1994projector, kresse1999ultrasoft}. $\Gamma$-centered 8$\times$8$\times$3 and 7$\times$7 Monkhorst-Pack meshes \cite{monkhorst1976special} were used to obtain wave-functions of the bulk compound and the monolayer, respectively.

\medskip
\textbf{Supporting Information} \par
Supporting Information is available from the Wiley Online Library or from the author.

\medskip
\textbf{Acknowledgements} \par
This work was supported by the Gordon and Betty Moore Foundation's EPIQS initiative through Grant GBMF9064, the David and Lucile Packard foundation, and the Sloan foundation. The authors acknowledge the use of Princeton’s Imaging and Analysis Center (IAC), which is partially supported by the Princeton Center for Complex Materials (PCCM), a National Science Foundation (NSF) Materials Research Science and Engineering Center (MRSEC; DMR-2011750). We acknowledge the support from the University of California Santa Barbara Quantum Foundry, funded by the NSF (DMR-1906325). The research reported here also made use of shared facilities of the UC Santa Barbara Materials Research Science and Engineering Center (NSF DMR-1720256). Y.M.O is supported by the NSF Graduate Research Fellowship Program under Grant No. DGE-1650114.

\medskip
\textbf{Conflict of interest} \par
The authors declare no conflict of interest.

\newpage

\textbf{Supplementary information: TaCo$_{2}$Te$_{2}$: An air-stable, magnetic van der Waals material with high mobility}\\

\setcounter{figure}{0}

Figure S1 shows the scanning electron microscopy (SEM) image of a TaCo$_{2}$Te$_{2}$ single crystal, illustrating the layered structure of the material.\\

\begin{figure*}
\includegraphics[width=0.7\textwidth]{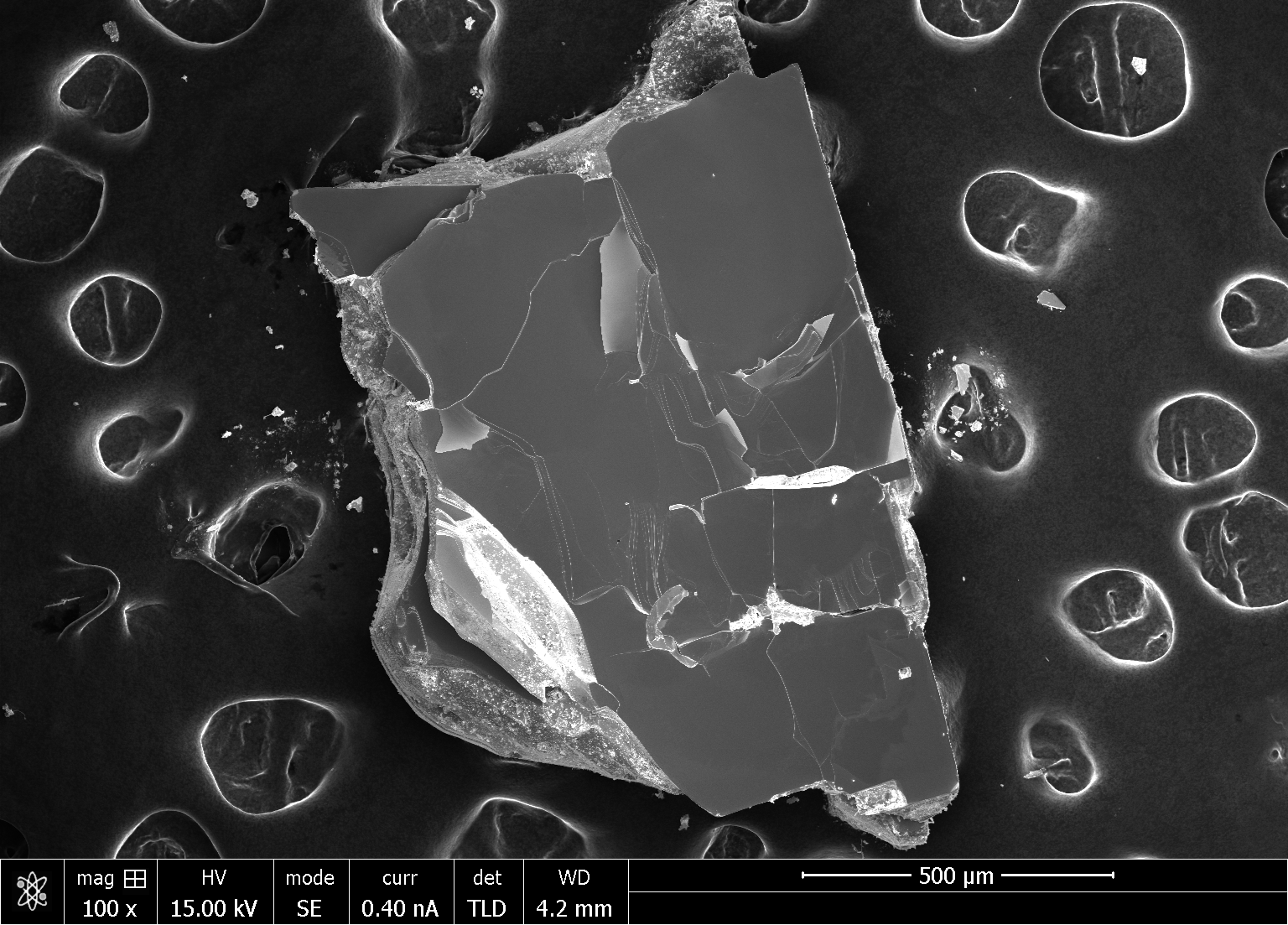}
\renewcommand{\figurename}{Fig.S}
\caption{Scanning electron microscopy image of a typical TaCo$_{2}$Te$_{2}$ single crystal.}
\end{figure*}

Figure S2 represents a typical energy dispersive x-ray spectroscopy (EDX) spectra of a grown crystal, confirming almost ideal chemical composition.\\

\begin{figure*}
\includegraphics[width=0.7\textwidth]{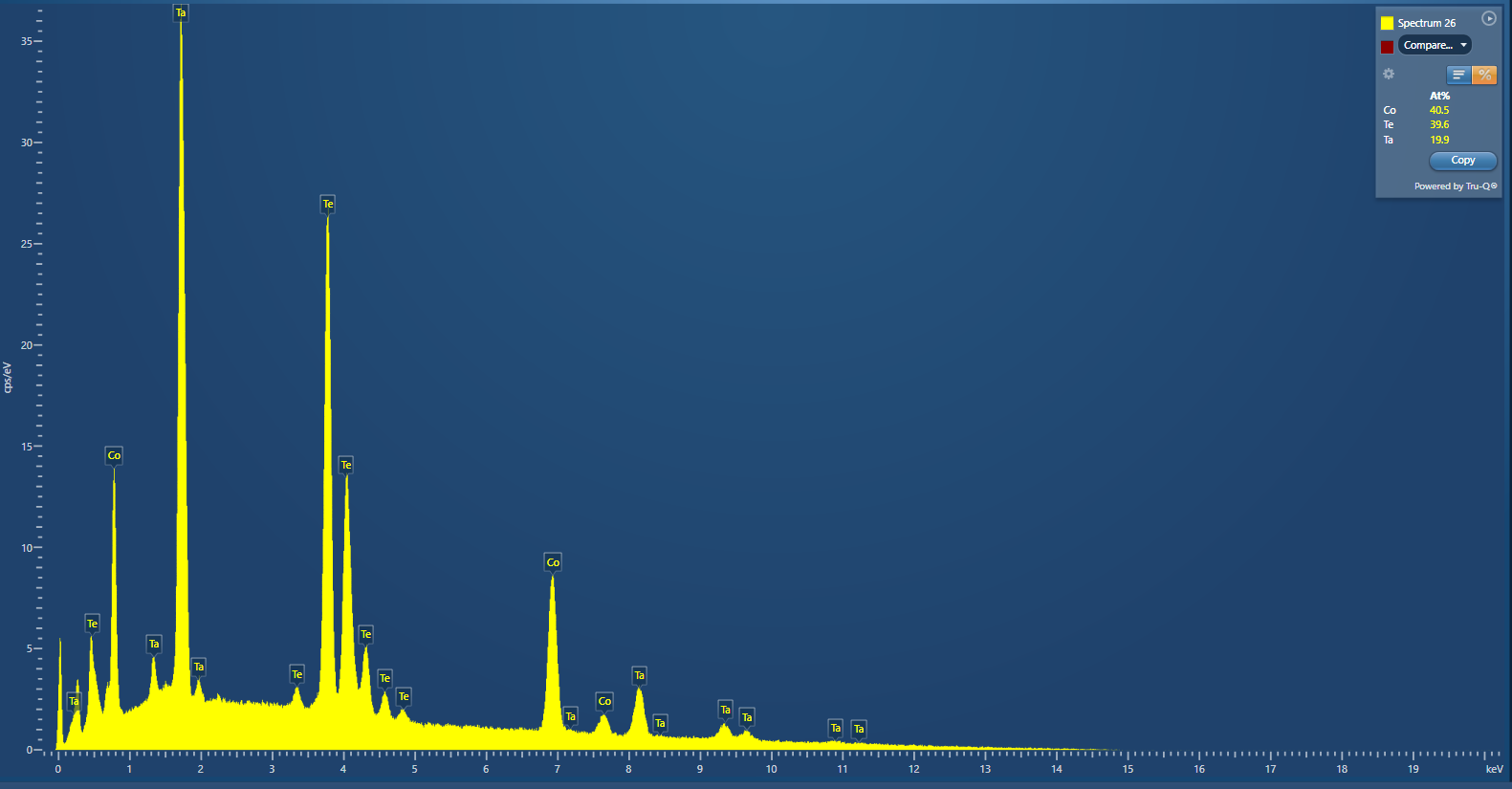}
\renewcommand{\figurename}{Fig.S}
\caption{A typical energy dispersive x-ray spectroscopy spectra for TaCo$_{2}$Te$_{2}$ single crystal.}
\end{figure*}

Table S1, S2, and S3 summarize the results of the structural solution, obtained from the single crystal x-ray diffraction measurements on the TaCo$_{2}$Te$_{2}$ crystals.\\

\begin{center}
\begin{table*}
\renewcommand{\tablename}{Table S}
\caption{Crystal data and structure refinement for TaCo$_{2}$Te$_{2}$.}
 \begin{tabular}{c c c c c c c c c c c c c c c c c c c c c c c c c c c c c c c c c c c c}
 \hline

  & Empirical formula & & & & & & & & & & & & & & & & & & & & & & & & & & & & & & & & & TaCo$_{2}$Te$_{2}$ &\\

  & Formula weight & & & & & & & & & & & & & & & & & & & & & & & & & & & & & & & & & 554.01 &\\

  & Wavelength & & & & & & & & & & & & & & & & & & & & & & & & & & & & & & & & & 0.71073 {\AA} &\\

  & Temperature & & & & & & & & & & & & & & & & & & & & & & & & & & & & & & & & & 295 K &\\

  & Crystal system & & & & & & & & & & & & & & & & & & & & & & & & & & & & & & & & & Orthorhombic &\\

  & Space group & & & & & & & & & & & & & & & & & & & & & & & & & & & & & & & & & $Pmcn$ (62) &\\

  & Unit cell dimensions & & & & & & & & & & & & & & & & & & & & & & & & & & & & & & & & & a=6.6015(4) {\AA}, $\alpha$=90$^{\circ}$ &\\

  & & & & & & & & & & & & & & & & & & & & & & & & & & & & & & & & & & b=6.5706(4) {\AA}, $\beta$=90$^{\circ}$ &\\

  & & & & & & & & & & & & & & & & & & & & & & & & & & & & & & & & & & c=17.7678(11) {\AA}, $\gamma$=90$^{\circ}$ &\\

  & Volume & & & & & & & & & & & & & & & & & & & & & & & & & & & & & & & & & 770.69(8) {\AA}$^{3}$ &\\

  & $Z$ & & & & & & & & & & & & & & & & & & & & & & & & & & & & & & & & & 8 &\\

  & Density (calculated) & & & & & & & & & & & & & & & & & & & & & & & & & & & & & & & & & 9.549 g/cm$^{3}$ &\\

  & Absorption coefficient & & & & & & & & & & & & & & & & & & & & & & & & & & & & & & & & & 51.424 mm$^{-1}$ &\\

  & $F$(000) & & & & & & & & & & & & & & & & & & & & & & & & & & & & & & & & & 1848 &\\

  & $\theta$ range for data collection & & & & & & & & & & & & & & & & & & & & & & & & & & & & & & & & & 2.293 to 29.989$^{\circ}$ &\\

  & Index ranges & & & & & & & & & & & & & & & & & & & & & & & & & & & & & & & & & 0$\leq h \leq$24, -9$\leq k \leq$9, -9$\leq l \leq$9 &\\

  & Reflections collected & & & & & & & & & & & & & & & & & & & & & & & & & & & & & & & & & 4259 &\\

  & Independent reflections & & & & & & & & & & & & & & & & & & & & & & & & & & & & & & & & & 1221 [$R_{int}$=0.0410] &\\

  & Completeness to $\theta$=25.242$^{\circ}$ & & & & & & & & & & & & & & & & & & & & & & & & & & & & & & & & & 100 \% &\\

  & Refinement method & & & & & & & & & & & & & & & & & & & & & & & & & & & & & & & & & Full-matrix least-squares on $F^{2}$ &\\

  & Data/restraints/parameters & & & & & & & & & & & & & & & & & & & & & & & & & & & & & & & & & 1221/0/52 &\\

  & Goodness-of-fit & & & & & & & & & & & & & & & & & & & & & & & & & & & & & & & & & 1.164 &\\

  & Final $R$ indices [$I>$2$\sigma$($I$)] & & & & & & & & & & & & & & & & & & & & & & & & & & & & & & & & & $R_{obs}$=0.0496, $wR_{obs}$=0.1021 &\\

  & $R$ indices [all data] & & & & & & & & & & & & & & & & & & & & & & & & & & & & & & & & & $R_{all}$=0.0606, $wR_{all}$=0.1068 &\\

  & Largest diff. peak and hole & & & & & & & & & & & & & & & & & & & & & & & & & & & & & & & & & 4.827 and -4.513 e{\AA}$^{-3}$ &\\

 \hline
\end{tabular}

$R=\Sigma||F_{0}|-|F_{c}||/\Sigma|F_{0}|$, $wR=\{\Sigma[w(|F_{0}|^{2}-|F_{c}|^{2})^{2}]/\Sigma[w(|F_{0}|^{4})]\}^{1/2}$,

and $w=1/[\sigma^{2}(F_{0})^{2}+(0.0258P)^{2}+80.6280P]$, where $P=(F_{0}^{2}+2F_{c}^{2})/3$.

\end{table*}
\end{center}

\begin{center}
\begin{table*}
\renewcommand{\tablename}{Table S}
\caption{Atomic coordinates ($\times$10$^{4}$) and equivalent isotropic displacement parameters ({\AA}$^{2}\times$10$^{3}$) for TaCo$_{2}$Te$_{2}$ at 295 K with estimated standard deviations in parentheses.}
 \begin{tabular}{c c c c c c c c c c c c c c c c c c c c c c c c c c c c c c c c c c c c c c c}
 \hline

 & Label & & & & & & & $x$ & & & & & & & $y$ & & & & & & & $z$ & & & & & & & Occupancy & & & & & & & $^{\ast}U_{eq}$ & &\\
 \hline

  & Ta(01) & & & & & & & 7500 & & & & & & & 6512(2) & & & & & & & 2652(1) & & & & & & & 1 & & & & & & & 7(1) & &\\

  & Ta(02) & & & & & & & 7500 & & & & & & & 1505(2) & & & & & & & 2219(1) & & & & & & & 1 & & & & & & & 8(1) & &\\

  & Te(03) & & & & & & & 2500 & & & & & & & 1533(2) & & & & & & & 3932(1) & & & & & & & 1 & & & & & & & 10(1) & &\\

  & Te(04) & & & & & & & 5030(2) & & & & & & & 6508(2) & & & & & & & 4062(1) & & & & & & & 1 & & & & & & & 10(1) & &\\

  & Te(05) & & & & & & & 7500 & & & & & & & 1496(2) & & & & & & & 4132(1) & & & & & & & 1 & & & & & & & 10(1) & &\\

  & Co(06) & & & & & & & 5328(3) & & & & & & & -425(3) & & & & & & & 3188(2) & & & & & & & 1 & & & & & & & 8(1) & &\\

  & Co(07) & & & & & & & 5342(3) & & & & & & & 3459(3) & & & & & & & 3190(1) & & & & & & & 1 & & & & & & & 8(1) & &\\

 \hline
\end{tabular}

$^{\ast}U_{eq}$ is defined as one third of the trace of the orthogonalized $U_{ij}$ tensor.

\end{table*}
\end{center}

\begin{center}
\begin{table*}
\renewcommand{\tablename}{Table S}
\caption{Anisotropic displacement parameters ({\AA}$^{2}$) for TaCo$_{2}$Te$_{2}$ at 295 K.}
 \begin{tabular}{c c c c c c c c c c c c c c c c c c c c c c c c c c c c c c c c c c c c c c c c c c c c}
 \hline

 & Label & & & & & & & $U_{11}$ & & & & & & & $U_{22}$ & & & & & & & $U_{33}$ & & & & & & & $U_{12}$ & & & & & & & $U_{13}$ & & & & & & & $U_{23}$\\
 \hline

  & Ta(01) & & & & & & & 0.00680 & & & & & & & 0.00660 & & & & & & & 0.00880 & & & & & & & 0.00000 & & & & & & & 0.00000 & & & & & & & -0.00010\\

  & Ta(02) & & & & & & & 0.00720 & & & & & & & 0.00850 & & & & & & & 0.00960 & & & & & & & 0.00000 & & & & & & & 0.00000 & & & & & & & 0.00020\\

  & Te(03) & & & & & & & 0.00850 & & & & & & & 0.01120 & & & & & & & 0.01040 & & & & & & & 0.00000 & & & & & & & 0.00000 & & & & & & & 0.00060\\

  & Te(04) & & & & & & & 0.01120 & & & & & & & 0.00830 & & & & & & & 0.00930 & & & & & & & -0.00050 & & & & & & & -0.00010 & & & & & & & 0.00010\\

  & Te(05) & & & & & & & 0.00860 & & & & & & & 0.00930 & & & & & & & 0.01060 & & & & & & & 0.00000 & & & & & & & 0.00000 & & & & & & & -0.00070\\

  & Co(06) & & & & & & & 0.00730 & & & & & & & 0.00660 & & & & & & & 0.01060 & & & & & & & 0.00020 & & & & & & & 0.00090 & & & & & & & 0.00020\\

  & Co(07) & & & & & & & 0.00870 & & & & & & & 0.00670 & & & & & & & 0.00850 & & & & & & & 0.00100 & & & & & & & -0.00090 & & & & & & & 0.00060\\

 \hline
\end{tabular}

The anisotropic displacement factor exponent takes the form: -2$\pi^{2}$[$h^{2}a^{\ast2}U_{11}+ ... + 2hka^{\ast}b^{\ast}U_{12}$].

\end{table*}
\end{center}

\begin{figure*}
\includegraphics[width=0.7\textwidth]{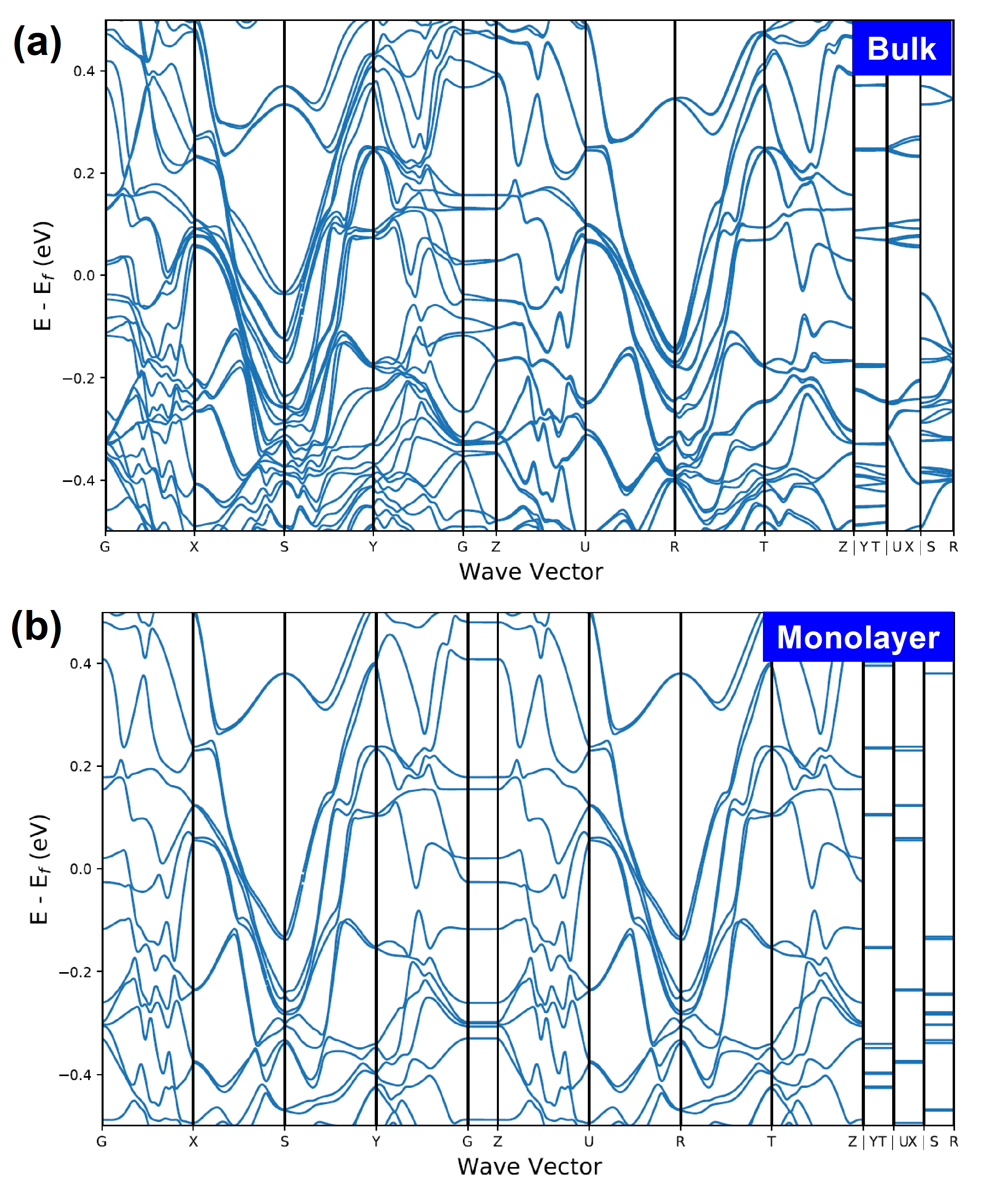}
\renewcommand{\figurename}{Fig.S}
\caption{Electronic band structure of TaCo$_{2}$Te$_{2}$ for a) the bulk compound and b) a monolayer.}
\end{figure*}

The results of the electronic band structure calculations are shown in Figure S3 for both the bulk compound and a monolayer.\\

\end{document}